\documentclass[aip,rsi,reprint,graphicx]{revtex4-1}
\usepackage{graphicx}  
\usepackage{dcolumn}   
\usepackage{bm}           
\usepackage{amssymb} 
\usepackage{amsmath} 

\begin{document}

\title{An open source digital servo for AMO physics experiments}

\author{D.~R.~Leibrandt}
\email{david.leibrandt@nist.gov}
\author{J.~Heidecker}
\affiliation{National Institute of Standards and Technology, Boulder, Colorado 80305, USA}

\date{\today}

\begin{abstract}
We describe a general purpose digital servo optimized for feedback control of lasers in atomic, molecular, and optical (AMO) physics experiments.  The servo is capable of feedback bandwidths up to roughly 1~MHz (limited by the 320~ns total latency); loop filter shapes up to fifth order; multiple-input, multiple-output control; and automatic lock acquisition.  The configuration of the servo is controlled via a graphical user interface, which also provides a rudimentary software oscilloscope and tools for measurement of system transfer functions.  We illustrate the functionality of the digital servo by describing its use in two example scenarios: frequency control of the laser used to probe the narrow clock transition of $^{27}$Al$^+$ in an optical atomic clock, and length control of a cavity used for resonant frequency doubling of a laser.
\end{abstract}


\maketitle

\section{Introduction}

Control is ubiquitous in atomic, molecular, and optical (AMO) physics.  A typical experiment might include a dozen or more feedback controllers for tasks such as laser frequency, power, and phase stabilization \cite{Young1999,Sparkes2011}; temperature stabilization \cite{Barone1995}; and vibration isolation \cite{Hensley1999}.  Yet the control repertoire practiced by an average AMO physicist is often limited to simple proportional-integral-derivative (PID) feedback, unable to leverage the developments of the field of control theory over the past several decades \cite{Dutton1997,Bechhoefer2005}.

With the rapid development of hardware suitable for low-latency digital signal processing (DSP), such as field-programmable gate arrays (FPGAs) and digital signal processors, high-speed digital feedback controllers have become increasingly practical and increasingly common over the past decade.  The advantages of a digital approach include fast and easy (no soldering) reconfiguration of the feedback transfer function;  implementation of multiple-input, multiple-output (MIMO) transfer functions with shapes that go beyond PID \cite{Barone1995,Hensley1999,Stockton2002,Jacky2008,Dietrich2009}; the ability to detect whether the system is currently locked and to perform automatic lock acquisition \cite{Allard2004,Xu2011,Huang2014}; the integration of signal pre-processing such as digital lock-in amplifiers for generation of the error signal \cite{Labaziewicz2007}; and the integration of diagnostics for easy analysis of open- and closed-loop system performance \cite{Sparkes2011}.  While for many years digital controllers were limited to low bandwidth applications such as temperature controllers, modern hardware is capable of MHz bandwidth feedback control.  However, there is typically a tradeoff between ease of use and performance, with microcontrollers being easier to program and less expensive but limited to feedback bandwidths below about 100~kHz \cite{Allard2004,Dietrich2009,Huang2014}, and digital signal processors and FPGAs being capable of several MHz feedback bandwidths but more expensive and usually requiring knowledge of specialized, low-level programming languages \cite{Stockton2002,Labaziewicz2007,Jacky2008}.

We have developed a general purpose digital servo optimized for feedback control of lasers in AMO physics experiments.  The servo is based on a custom-built hardware box that includes an FPGA for computation of feedback transfer functions, two channels of low-noise and high-speed analog-to-digital conversion (ADC), and three channels of low-noise and high-speed digital-to-analog conversion (DAC).  The hardware latency is suitable for feedback to acousto-optic modulators (AOMs) with bandwidths of several hundred kHz, and one of the analog output channels includes a high-voltage amplifier for driving piezoelectric transducers (PZTs).  Configuration of the feedback transfer function and diagnostic tools are controlled via a graphical user interface (GUI) that runs on a standard personal computer (PC).  The hardware and software design is public domain and available for download online \cite{DigitalServoWebsite}.  Others are encouraged to use the digital servo as is (which does not require knowledge of any programming languages), or to modify the design to suit their own purposes and to contribute their modifications to the project website.  The goal of this project is to lower the entry barrier for the use of high-performance digital control tools, enabling physicists to go beyond simple PID.

This paper proceeds as follows.  Section~\ref{sec:design} describes the hardware, firmware, and software design of the digital servo.  Section~\ref{sec:applications} describes the use of the digital servo in two example applications: frequency control of the clock laser used in an $^{27}$Al$^+$ optical atomic clock, and length control of a cavity used for resonant frequency doubling of a laser.  Some of the advantages of the digital servo relative to analog servos are highlighted.  Section~\ref{sec:conclusion} summarizes and concludes.  Finally, Appendix~\ref{sec:DSPprimer} presents a detailed description of the digital filter design used in the digital servo.

\section{Design}\label{sec:design}

The digital servo is comprised of a custom-built hardware box that implements the desired feedback transfer function using infinite impulse response (IIR) filters in an FPGA, firmware that describes the DSP in the FPGA, and a software GUI that runs on a PC for setting the feedback transfer function and controlling diagnostic tools.  These three components will be described in the following three subsections.

\subsection{Hardware}

The digital servo hardware consists of two printed circuit boards (PCBs) packaged in a $56 \times 200 \times 205$~mm$^3$ box.  The first is a commercial FPGA integration module (Opal Kelly XEM6010-LX150 \cite{CommercialProduct}), which contains a Xilinx Spartan 6 XC6SLX150-2 FPGA \cite{CommercialProduct}, a 32 MB flash memory, a 128 MB SDRAM, and a universal serial bus (USB) 2.0 interface.  This FPGA was selected for its combination of high DSP performance and relatively low cost.  Specifically, it contains 180 DSP48A1 slices that are each capable of performing $18 \times 18$ bit signed integer multiplication in a single clock cycle.  The $35 \times 35$ bit signed integer multiplications used to compute the IIR filters in the digital servo are performed using four DSP48A1 slices combined to form a single multiplier.  Although the maximum clock rate for these $35 \times 35$ multipliers is specified to be 39~MHz in the timing report generated by the Xilinx ISE Design Suite \cite{CommercialProduct}, we overclock them at 100~MHz without any pipeline stages and have not seen any resulting multiplication errors.  Note, however, that all testing and operation of the digital servos reported here took place in well temperature controlled laboratories, and that temperatures or supply voltages closer to the specified limits of the FPGA may cause multiplication errors at this clock rate.  The flash memory is used to store both the FPGA firmware and the servo feedback configuration while the power is off.  The SDRAM is used to store the values of the digitized signals at sample rates up to 6~MHz; this data is read back to the PC at a slower rate for off-line analysis of noise spectra and transfer functions.  The digital servo hardware communicates with the PC over the USB interface.

The second PCB is a custom-built daughterboard that provides two channels of high-speed analog input, two channels of high-speed analog output, and one channel of low-speed analog output.  The high-speed inputs and outputs are implemented by a two channel, 16 bit ADC (Linear Technology LTC2195 \cite{CommercialProduct}) and a two channel, 16 bit DAC (Analog Devices AD9783 \cite{CommercialProduct}), both of which operate with update rates of 100~MHz.  These chips were selected for their low pipeline delays, their low noise, and their use of low voltage differential signal (LVDS) interfaces.  Input and output range scaling of the high-speed channels is provided by variable gain amplifiers (VGAs, Analog Devices AD8251 \cite{CommercialProduct}), so that the high-speed inputs have a software selectable range of $\pm 0.5$~V, $\pm 1$~V, $\pm 2$~V, or $\pm 4$~V and the high-speed outputs have a software selectable range of $\pm 1$~V, $\pm 2$~V, $\pm 4$~V, or $\pm 8$~V.  The analog bandwidth of the high speed inputs and outputs is limited by the VGAs to between 3~MHz and 10~MHz depending on the gain setting.  The low-speed output is implemented by a single channel, 20 bit DAC (Analog Devices AD5791 \cite{CommercialProduct}) that operates with a 1~MHz update rate.  Two output amplifiers are included so that the output range 0~V to $+10$~V is available on one output connector and the output range 0~V to $+65$~V is available on a second output connector.  The daughterboard also provides several channels of digital input, which can be used for integrator hold or automatic lock acquisition functionality, and digital output, which can be used to tell other parts of a complex experiment whether the servo is currently locked.

The input noise of the analog inputs when set to the $\pm 0.5$~V input range and shorted to ground is 50~nV$/\sqrt{\textrm{Hz}}$ at 1~kHz, and the long-time stability is 2~$\mu$V for averaging times between $10^{-3}$~s and $10^3$~s (in a well temperature controlled laboratory).  Note that this stability is smaller than one bit of the 16 bit ADC, which is possible because we are oversampling in the regime where the analog noise at the input of the ADC is slightly larger than one bit \cite{Stewart1998}.  This dithering is important as it suppresses limit cycle oscillations.  The output noise of the high-speed analog outputs when set to the $\pm 1$~V output range is 40~nV$/\sqrt{\textrm{Hz}}$ at 1~kHz.  The output noise of the low-speed analog output is 20~nV$/\sqrt{\textrm{Hz}}$ at the 0~V to $+10$~V output range connector and 300~nV$/\sqrt{\textrm{Hz}}$ at the 0~V to $+65$~V output range connector at 1~kHz.

\subsection{Firmware}\label{sec:firmware}

\begin{figure*}
\begin{center}
\includegraphics[width=1.0\textwidth]{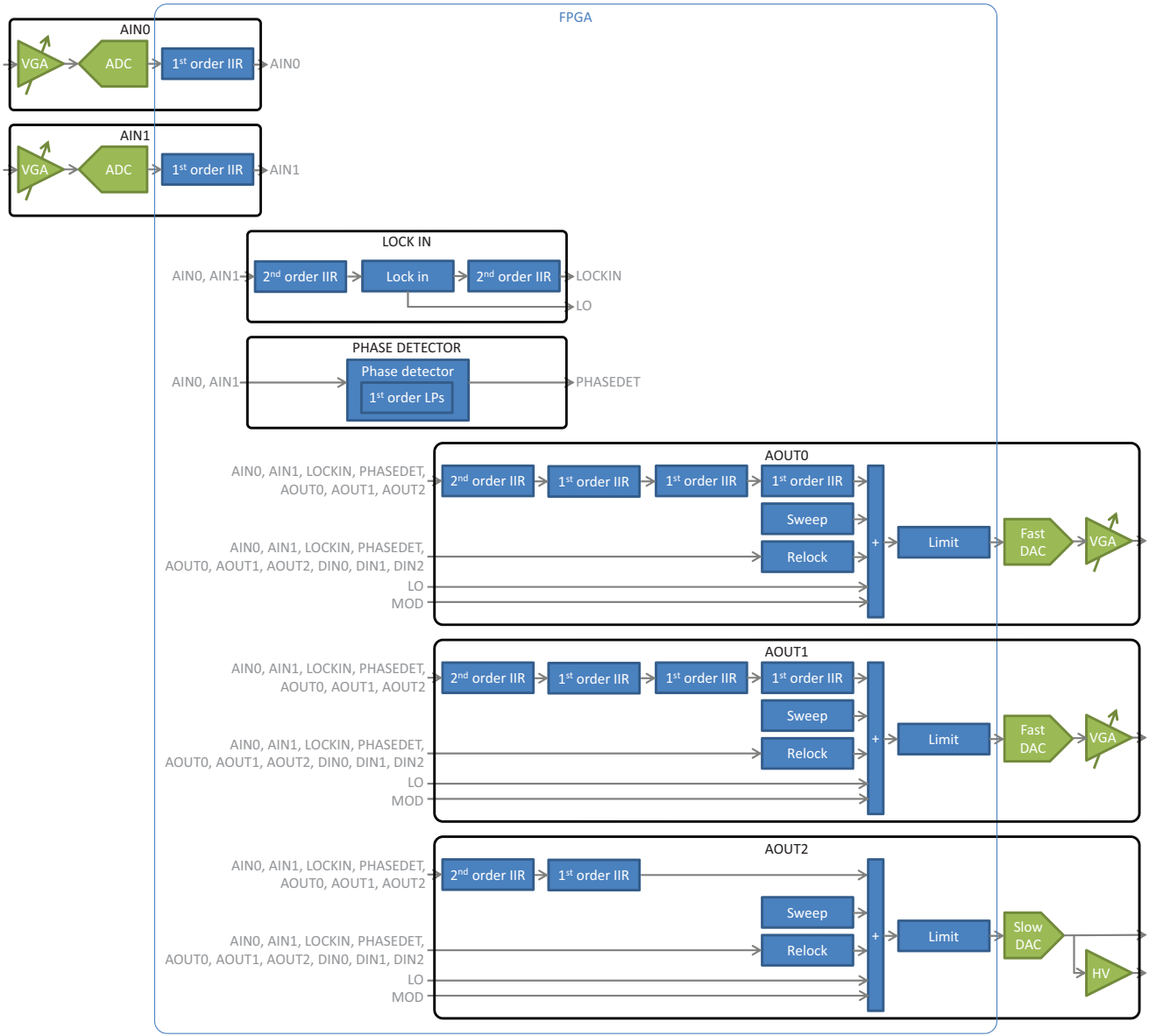}
\caption{\label{fig:DigitalServoBlockDiagram}Block diagram depicting the digital servo DSP modules and signal paths.  Green blocks are analog components and blue blocks are DSP modules implemented in the FPGA.  Note that this FPGA configuration uses 164 out of 180 available DSP48A1 slices.  Utilization of all other types of FPGA resources is below 50\%.  FPGA: field-programmable gate array; VGA: variable gain amplifier; ADC: analog-to-digital converter; DAC: digital-to-analog converter; IIR: infinite impulse response filter; LP: low-pass filter; LO: lock-in local oscillator; MOD: modulation signal used for transfer function measurement; AINx: analog input channel x; AOUTx: analog output channel x; DINx: digital input channel x; DOUTx: digital output channel x.}
\end{center}
\end{figure*}

The digital servo firmware is written in the Verilog hardware description language \cite{Palnitkar2003} using the Xilinx ISE Design Suite \cite{CommercialProduct}.  Figure~\ref{fig:DigitalServoBlockDiagram} shows a block diagram depicting the DSP modules and signal paths.  Most of the calculations are clocked at 100~MHz.  The discretized analog signals are passed between DSP modules in the FPGA as 24 bit signed integer signals, and the IIR filters internally use 35 bit signed integer signals to represent both the data and the coefficients to prevent rounding errors \cite{Jacky2008}.  The settings of each module and all of the signal connections not pictured in Fig.~\ref{fig:DigitalServoBlockDiagram} are software configurable in the GUI, and many of the IIR filters can be optionally bypassed when they are not needed to minimize the latency.

Each of the analog inputs is immediately followed by a first-order IIR filter.   This filter can be either enabled or disabled (bypassed) and is typically configured as a low-pass filter to remove input noise above the desired feedback bandwidth.

Each analog output signal is generated as the sum of a loop filter module (i.e., feedback controller), a sweep module that generates a triangle wave, and a relock module that performs automatic lock acquisition.  The loop filter module can take any of the analog signals as its input and consists of up to four sequentially applied first- or second-order IIR filters.  For example, a proportional-integral-integral-derivative (PIID) filter, which is third order, can be configured by enabling three sequential first order IIR filters (PI, PI, PD where PI is a proportional-integral filter and PD is a proportional-derivative filter).  For a comprehensive description of the IIR filters implemented in the digital servo as well as Bode diagrams of the built-in IIR filter transfer functions, see Appendix~\ref{sec:DSPprimer}.  The relock module can accept any of the analog signals or digital inputs as its input.  If the input signal is within the software defined ``locked'' range, the relock module output is zero.  If the input signal falls outside the ``locked'' range, the relock module holds the output (and internal state) of the loop filter module constant and outputs a triangle wave sweep with an amplitude that doubles every cycle and a constant slew rate.  Once the input signal returns to the ``locked'' range, the loop filter module is re-enabled and the relock module output ramps back to zero.

\begin{table*}
\caption{\label{tab:DSPmodules}Computational cost of DSP modules and hardware used in the digital servo.  Note that the latency and throughput values assume a 100~MHz FPGA clock.  Communication with both the ADC and the high-speed DAC is a combination of parallel and serial, and the firmware latencies are dominated by serialization and deserialization of the data.}
\begin{ruledtabular}
\begin{tabular}{llll}
Module							& Latency [ns]	& Throughput [MHz]& DSP48A1 slices \\
\hline
ADC total							& 130			& 100			& 0 \\
\hspace{4ex}hardware					& 88			& --			& -- \\
\hspace{4ex}firmware					& 42			& --			& -- \\
High-speed DAC total					& 90			& 100			& 0 \\
\hspace{4ex}hardware					& 70			& --			& -- \\
\hspace{4ex}firmware					& 20			& --			& -- \\
Low-speed DAC total					& 1330		& 1			& 0 \\
\hspace{4ex}hardware					& 180			& --			& -- \\
\hspace{4ex}firmware					& 1150		& --			& -- \\
First order low-pass IIR filter				& 30			& 100			& 8 \\
First order generic IIR filter				& 30			& 100			& 12 \\
Second order generic IIR filter (time multiplexed)	& 80			& 100/27		& 4 \\
Lock-in amplifier (not including pre and post filters)	& 10			& 100			& 6 \\
Phase detector (including post filters)			& 240			& 100			& 25 \\
Transfer function measurement				& --			& 100			& 5 \\
\end{tabular}
\end{ruledtabular}
\end{table*}

\begin{figure}
\begin{center}
\includegraphics[width=1.0\columnwidth]{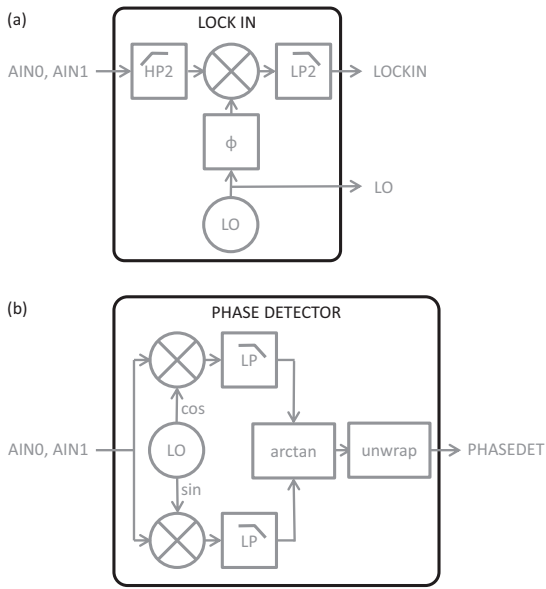}
\caption{\label{fig:LockInAndPhaseDetBlockDiagram}Block diagram showing details of the lock-in amplifier and phase detector modules. AINx: analog input channel x; HP2: second-order high-pass IIR filter; LP2: second-order low-pass IIR filter; $\phi$: phase shifter; LO: local-oscillator; LP: first-order low-pass IIR filter; LOCKIN: lock-in amplifier module output signal; PHASEDET: phase detector module output signal.}
\end{center}
\end{figure}

In addition to the input and output signals discussed above, there are digital lock-in amplifier and digital phase detector modules implemented in the FPGA (see Fig.~\ref{fig:LockInAndPhaseDetBlockDiagram}).  The lock-in module takes either of the analog inputs, optionally applies a pre-filter (typically a high-pass filter), multiplies it by an internally generated local oscillator, and optionally applies a post-filter (typically a low-pass filter).  The lock-in local oscillator (LO) signal can be summed onto any of the analog outputs.  The phase detector module takes either of the analog inputs, multiplies it by both quadratures of a second internally generated LO, low-pass filters the products, and then takes the arctangent of the ratio of the results.  The phase detector LO can optionally be clocked from an external 10~MHz reference.  The phase detector output is unwrapped such that it can track phase noise with an amplitude up to $\pm 2^{16} \pi$ radians.  Undetected cycle slips are avoided as long as the absolute value of the input phase minus the LO phase changes by less than $\pi$ radians in each 10~ns FPGA clock cycle.

The minimum latency of the servo, measured as the time delay for a signal to propagate from an analog input to one of the fast analog outputs, is $\tau = 320$~ns including a single first-order IIR filter.  This limits the feedback bandwidth to roughly 1~MHz (more precisely there is a $\pi$ phase shift at $f = 1/(2 \tau) = 1.6$~MHz).  Additional IIR filters and other functionality can be enabled at the cost of increased latency and hence reduced feedback bandwidth.  Table~\ref{tab:DSPmodules} lists the latency of several of the DSP modules used in the digital servo.  The latency of each module should be kept in mind when selecting a loop filter configuration for high-bandwidth applications.

\subsection{Software}

The digital servo software is written in the C++ programming language and uses the Qt application development framework \cite{Blanchette2008} for creation of the GUI.  Qt enables easy portability between different operating systems, and we have compiled and run the digital servo GUI on both Microsoft Windows \cite{CommercialProduct} and Linux.

The primary functionality of the GUI is to provide (virtual) knobs for setting the feedback transfer functions.  Internally, the GUI controls the configuration and the connectivity of the DSP modules described in Sec.~\ref{sec:firmware}.  Additionally, the GUI includes a rudimentary software oscilloscope that can be used to monitor the performance of the system and controls for recording the values of the digital servo signals to files on the PC.  Finally, the GUI can implement slow feedback to control the temperature of lasers via a recommended standard 232 (RS-232) serial port.

The digital servo hardware and firmware are capable of running autonomously, without being connected to a PC.  Of course, most of the configuration settings cannot be changed in this mode, but we have found that for some applications the relock module is able to keep the system locked without any human intervention for weeks at a time.  This capability also allows the digital servo to run continuously while the control PC is restarted.

\section{Experimental applications}\label{sec:applications}

This section presents two example applications of the digital servo, which serve to illustrate some of the advantages over an analog servo.  The first is a frequency servo for a laser that is used to drive the narrow $^1$S$_0$~$\leftrightarrow$~$^3$P$_0$ transition of $^{27}$Al$^+$ in an optical atomic clock \cite{Chou2010a}.  This example illustrates MIMO feedback and the automatic relocking feature of the digital servo.  The second is a length servo for a laser frequency doubling cavity \cite{Wilson2011}.  This example illustrates use of the digital servo for system transfer function measurement and implementation of a feedback transfer function which includes a notch filter.

\subsection{Clock laser frequency servo}\label{sec:clock-laser}

\begin{figure}
\begin{center}
\includegraphics[width=0.85\columnwidth]{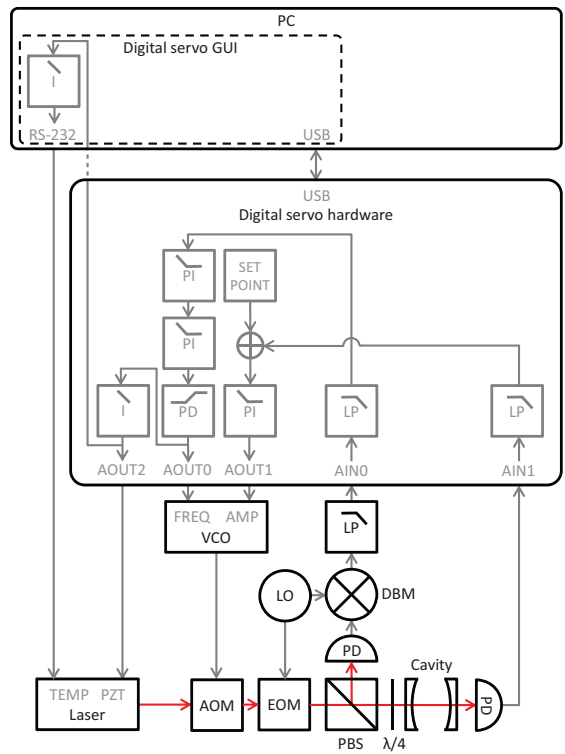}
\caption{\label{fig:ClockLaserBlockDiagram}Block diagram depicting feedback control of the laser frequency of a laser used for interrogation of the clock transition of $^{27}$Al$^+$.  Gray paths denote electrical signal propagation, and red paths denote laser propagation.  RS-232: recommended standard 232 serial port; USB: universal serial bus; I: integral IIR filter; PI: proportional-integral IIR filter; PD: (gain limited) proportional-derivative IIR filter; LP: low-pass IIR filter; FREQ: frequency tuning port; AMP: amplitude tuning port; VCO: voltage-controlled oscillator; LP: low-pass filter; LO: local oscillator; DBM: double-balanced mixer; PD: photodiode; TEMP: temperature tuning port; PZT: piezoelectric transducer tuning port; AOM: acousto-optic modulator; EOM: electro-optic phase modulator; PBS: polarizing beam splitter; $\lambda/4$: quarter-wave plate.}
\end{center}
\end{figure}

We have used the digital servo to lock the frequency of a 1070~nm fiber laser to a Fabry-P\'erot cavity, as described by \citet{Leibrandt2013} and shown schematically in Fig.~\ref{fig:ClockLaserBlockDiagram}.  The cavity length is thermally shifted by the circulating laser power in the cavity, so we must also stabilize the circulating power.  The actuators in this example are the temperature of the laser (for slow laser frequency corrections), a PZT in the laser (for intermediate speed laser frequency corrections), the frequency of a voltage-controlled oscillator (VCO) that drives a fiber-coupled acousto-optic modulator (AOM, for fast laser frequency corrections), and the amplitude of the VCO (for laser power corrections).  The laser frequency error signal is derived from the cavity reflection by the Pound-Drever-Hall (PDH) method \cite{Drever1983}, and the circulating power error signal is generated by a photodiode which monitors the laser power transmitted through the cavity.  The transfer function for fast laser frequency feedback via the VCO frequency is PIID.  The derivative gain is used to increase the achievable feedback bandwidth, which is limited to 500~kHz by the acoustic-wave propagation time in the AOM (i.e., a time delay).  Since the frequency tuning range of the AOM is limited to a few MHz, we also feed-back to the laser PZT with a bandwidth of the order of 10~Hz and a frequency tuning range of a few tens of MHz, and we feed-back to the laser temperature with a bandwidth of the order of 10~mHz and an unlimited frequency tuning range.  All of these feedback paths are implemented using a single digital servo hardware box.  The AOM frequency feedback bandwidth is sufficient to significantly reduce the laser linewidth from its unlocked value of roughly 5~kHz.  The fractional frequency stability of this laser when locked to the Fabry-P\'erot cavity is $2 \times 10^{-15}$ for averaging times between 0.5~s and 10~s \cite{Leibrandt2013}, limited by thermomechanical length fluctuations of the cavity \cite{Numata2004} rather than servo noise.

\begin{figure}
\begin{center}
\includegraphics[width=1.0\columnwidth]{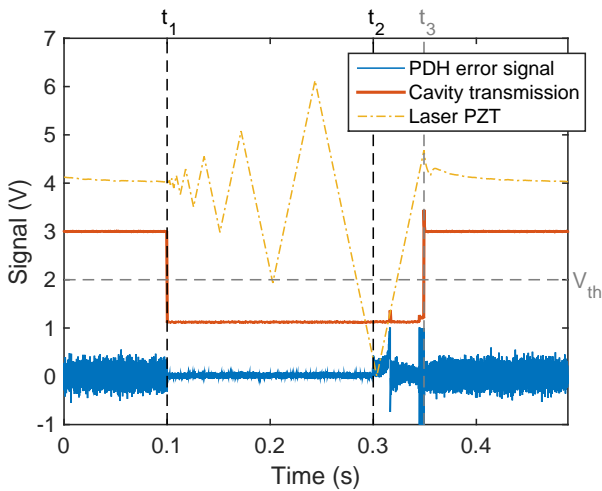}
\caption{\label{fig:ClockLaserRelock}Automatic relocking of the clock laser.  The digital servo analog input and output signals are plotted as a function of time; the cavity transmission trace is scaled and offset by $+1$~V for clarity.  The laser power is switched off between $t_1 = 0.1$~s and $t_2 = 0.3$~s.  The digital servo enables the servo at $t_3 = 0.349$~s when the cavity transmission crosses the threshold $V_{th} = 2$~V.}
\end{center}
\end{figure}

The digital servo automatically acquires lock to the cavity upon turning the power on, and automatically relocks when a disturbance causes the laser to fall out of lock.  Figure~\ref{fig:ClockLaserRelock} shows the behavior of the digital servo relock module when the laser power is briefly switched off using an AOM.  The servo determines if the laser is locked to the cavity by checking if the transmitted power is greater than a threshold set in the GUI.  When the laser power is switched off at $t_1 = 0.1$~s, the outputs of the laser temperature and PZT loop filters are held constant and the PZT is swept with a constant slew rate and a gradually increasing amplitude.  The laser power is switched back on at $t_2 = 0.3$~s.  Note that the threshold is set such that when the 10~MHz PDH phase modulation sideband of the laser sweeps over the cavity resonance at 0.316~s, the servos are not engaged.  Finally, at $t_3 = 0.349$~s the laser phase modulation carrier sweeps over the cavity resonance and the loop filters are turned back on.  Since the relock functionality of the digital servo is implemented in the FPGA, it can be much faster than is shown here, limited only by the bandwidth and latency of the analog inputs and outputs.

\subsection{Laser frequency doubling cavity servo}

\begin{figure}
\begin{center}
\includegraphics[width=1.0\columnwidth]{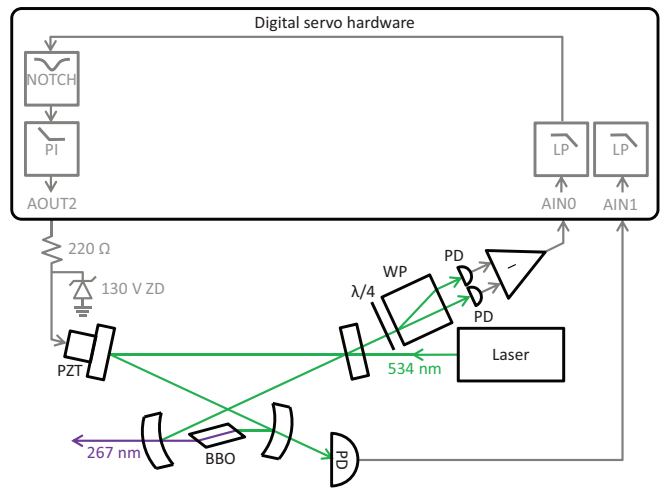}
\caption{\label{fig:DoublingCavityBlockDiagram}Block diagram depicting feedback control used to lock a laser frequency doubling cavity to a laser.  Gray paths denote electrical signal propagation, green paths denote visible laser light, and purple paths denote ultraviolet laser light.  Note that the Zener diode is used to protect the piezoelectric transducer.  NOTCH: notch IIR filter; PI: proportional-integral IIR filter; LP: low-pass IIR filter; ZD: Zener diode; PZT: piezoelectric transducer; $\lambda/4$: quarter-wave plate; WP: Wollaston prism; PD: photodiode; BBO: $\beta$-BaB$_2$O$_2$ nonlinear crystal.}
\end{center}
\end{figure}

We have also used the digital servo to lock the resonance of a laser frequency doubling cavity to a laser, as shown schematically in Fig.~\ref{fig:DoublingCavityBlockDiagram}.  The doubling cavity implements nonlinear frequency conversion from 534~nm to 267~nm in a BBO ($\beta$-BaB$_2$O$_2$) nonlinear crystal, and is based on the one used by \citet{Wilson2011}.  The actuator is a PZT that tunes the length of the cavity, and the error signal is derived from the cavity reflection using the H\"{a}nsch-Couillaud (HC) method \cite{Hansch1980}.  The 534~nm buildup light which leaks through one of the cavity mirrors is monitored by a photodiode and used for automatic relocking, similar to that described in Sec.~\ref{sec:clock-laser}.

We measure the transfer function of the doubling cavity PZT by adding a swept sine wave to the digital servo output and recording the resulting error signal at the digital servo input.  This functionality is only partially automated: the digital servo GUI automatically collects the data but the data processing happens manually off-line.  An alternative approach has been demonstrated by \citet{Sparkes2011} in which the transfer function is measured at all frequencies simultaneously using white noise.

\begin{table}
\caption{\label{tab:SISOtransferFunction}Summary of measurements used to determine the transfer function of a SISO plant while feedback is applied to keep the plant near the setpoint.}
\begin{ruledtabular}
{\renewcommand{\arraystretch}{2.0}
\begin{tabular}{llll}
\multicolumn{2}{l}{Configuration}	& Measured									& Inferred \\
$G$	& $K$				& transfer function								& transfer function \\
\hline
1	& 0				& $\frac{z}{m} \equiv X_{AD} = A D$					& $A D = X_{AD}$ \\
1	& $K$				& $\frac{z}{m} \equiv X_{K} = \frac{A D}{1 + A D K}$		& $K = \frac{1}{X_{K}} - \frac{1}{AD}$ \\
$G$	& $K$				& $\frac{z}{m} \equiv X_{G} = \frac{A D G}{1 + A D G K}$		& $G = \frac{X_{G}}{AD (1 - K X_{G})}$ \\
\end{tabular}
}
\end{ruledtabular}
\end{table}

\begin{figure}
\begin{center}
\includegraphics[width=1.0\columnwidth]{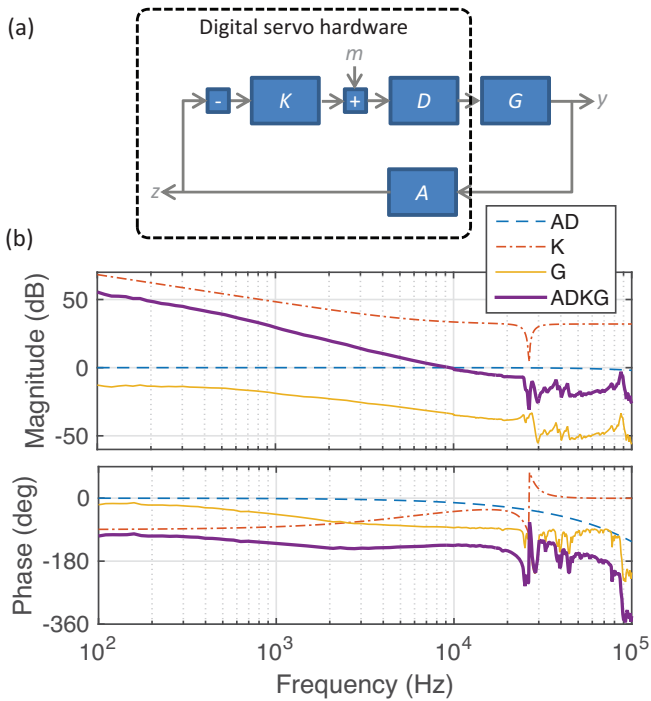}
\caption{\label{fig:DoublingCavityTransferFunction}(a) Block diagram illustrating use of the digital servo for measurement of the transfer function of a laser frequency doubling cavity.  K: controller (IIR filter implemented in the FPGA); m: modulation (added to the controller output for transfer function measurement); D: digital-to-analog converter; G: plant (laser frequency doubling cavity); y: plant output; A: analog-to-digital-converter; z: discretized plant output.  (b) Measured transfer functions of the digital servo DAC and ADC (AD, blue dashed line) and the doubling cavity PZT (G, yellow line, scaled by -26~dB for clarity), as well as calculated transfer functions of the digital servo (K, red dash dotted line, scaled by +26~dB for clarity) and the product of all of the above transfer functions (ADKG, thick purple line).  The corner frequency of the PI filter used in the feedback transfer function is 6500~Hz.}
\end{center}
\end{figure}

Since the HC method produces an error signal which is only linear for small deviations from the cavity resonance, we measure the transfer function while the cavity length is locked to the laser.  The block diagram of the digital servo controlling a generic single-input, single-output (SISO) plant is shown in Fig.~\ref{fig:DoublingCavityTransferFunction}(a).  This can be translated into an equation for the digitized output $z$ as a function of the modulation $m$:
\begin{equation}
z = \frac{A D G}{1 + A D G K} m \ .
\end{equation}
In order to determine the transfer function of the plant $G$ (i.e., the laser frequency doubling cavity), we measure the transfer function $z/m$ for two different configurations of the signal path summarized in Tab.~\ref{tab:SISOtransferFunction}.  In the first measurement (Tab.~\ref{tab:SISOtransferFunction}, line 1), the output of the digital servo is connected directly to the input of the digital servo, bypassing the plant altogether ($G \rightarrow 1$), and the loop filter is turned off ($K \rightarrow 0$).  This measurement gives us the product $A D$, which is the combined transfer function of the analog-to-digital converter and the digital-to-analog converter.  In the second measurement (Tab.~\ref{tab:SISOtransferFunction}, line 3), the digital servo is connected to the plant ($G \rightarrow G$) and the loop filter is turned on ($K \rightarrow K$).  This measurement, combined with the previous measurement and a priori knowledge of the transfer function of the loop filter $K$, gives us the transfer function of the plant $G$.  Note that if the transfer function of the loop filter were not known a priori, it could be determined by a third measurement (Tab.~\ref{tab:SISOtransferFunction}, line 2).  The measured transfer function of the doubling cavity PZT is shown in Fig.~\ref{fig:DoublingCavityTransferFunction}(b).  Note that the 1~kHz low-pass component of the PZT transfer function is due to an electrical low-pass filter composed of the 220~$\Omega$ resistor shown in Fig.~\ref{fig:DoublingCavityBlockDiagram} and the 0.75~$\mu$F capacitance of the PZT, which is used to reduce the voltage noise of the digital servo analog output at high frequencies.

The doubling cavity PZT has several mechanical resonances, with the lowest being at 25~kHz.  We use a notch filter in the feedback transfer function to suppress oscillations at this frequency, which allows for an improved feedback bandwidth.  The transfer function $ADKG$ shown in Fig.~\ref{fig:DoublingCavityTransferFunction}(b) has a gain margin of 2.0 and a phase margin of 45~degrees, with a feedback bandwidth of 10~kHz.  In comparison, an analog PID filter can only achieve a 6 kHz feedback bandwidth with the same gain and phase margin for this doubling cavity, limited by the requirement that the loop gain be below unity at the PZT resonance frequencies.

\section{Conclusion}\label{sec:conclusion}

We have described a digital servo with two analog input channels and three analog output channels, capable of MIMO control with bandwidths up to roughly 1~MHz.  Computation of the feedback transfer function takes place in an FPGA for high-speed and deterministic timing, and the servo configuration is controlled via a GUI that runs on a PC.  The digital servo can implement feedback transfer functions consisting of up to four cascaded first- and second-order IIR filters, which allows for loop shapes such as PIID or PI with a notch filter, as well as automatic relocking.  Diagnostic functionality includes a rudimentary software oscilloscope and measurement of system transfer functions.  We presented two example applications of the digital servo that illustrate the above capabilities.  The hardware and software design is public domain and available for download online \cite{DigitalServoWebsite}, and others are encouraged to try it out in their own laboratories.

While the design is optimized for feedback control of lasers in AMO physics experiments, it is intended to be general purpose and it should be applicable to other control applications with similar bandwidth, noise, and loop shape requirements.  Furthermore, with modified firmware, it is possible to perform more specialized control tasks using the same hardware.  In addition to the general purpose digital servo described here, this hardware has also been used for low-drift digital demodulation of a PDH error signal \cite{Cook2015}, feedforward corrections of acceleration induced laser frequency noise \cite{Leibrandt2013}, and phase locking of a robust, portable frequency comb \cite{Sinclair2014}.

\begin{acknowledgements}
We thank Till Rosenband and Rahul Mhaskar for useful discussions, and Robert J\"{o}rdens and Chin-Wen Chou for careful reading of the manuscript. This paper as well as the hardware and software design of the digital servo are contributions of the U.S. Government, not subject to U.S.~copyright. This work is supported by the Defense Advanced Research Projects Agency (DARPA), the Office of Naval Research (ONR), and the Office of the Director of National Intelligence (ODNI) Intelligence Advanced Research Projects Activity (IARPA).
\end{acknowledgements}


\begin{thebibliography}{26}
\bibitem [{Com()}]{CommercialProduct}%
  \BibitemOpen
  \href@noop {} {\bibinfo  {journal} {Commercial products are identified for
  technical clarity. Such identification does not imply endorsement by the
  National Institute of Standards and Technology}\ }\BibitemShut {NoStop}%
\end{thebibliography}

%

\appendix
\section{Digital signal processing primer}\label{sec:DSPprimer}

The following is a description of the theory behind the DSP used in the digital servo.  For more details, please consult the control theory review article by \citet{Bechhoefer2005} or one of the many control theory textbooks (e.g., \citet{Dutton1997}).

Most of the DSP performed by the digital servo consists of first- and second-order IIR filters.  Higher-order filters are built up by sequential application of several first- and second-order filters.  This is better than using a single higher-order IIR filter both because it minimizes the discretization errors due to finite precision integer math and because the modularity of this approach makes filter design and programming easier.

\begin{figure*}
\begin{center}
$
\begin{array}{cc}
\includegraphics[width=1.0\columnwidth]{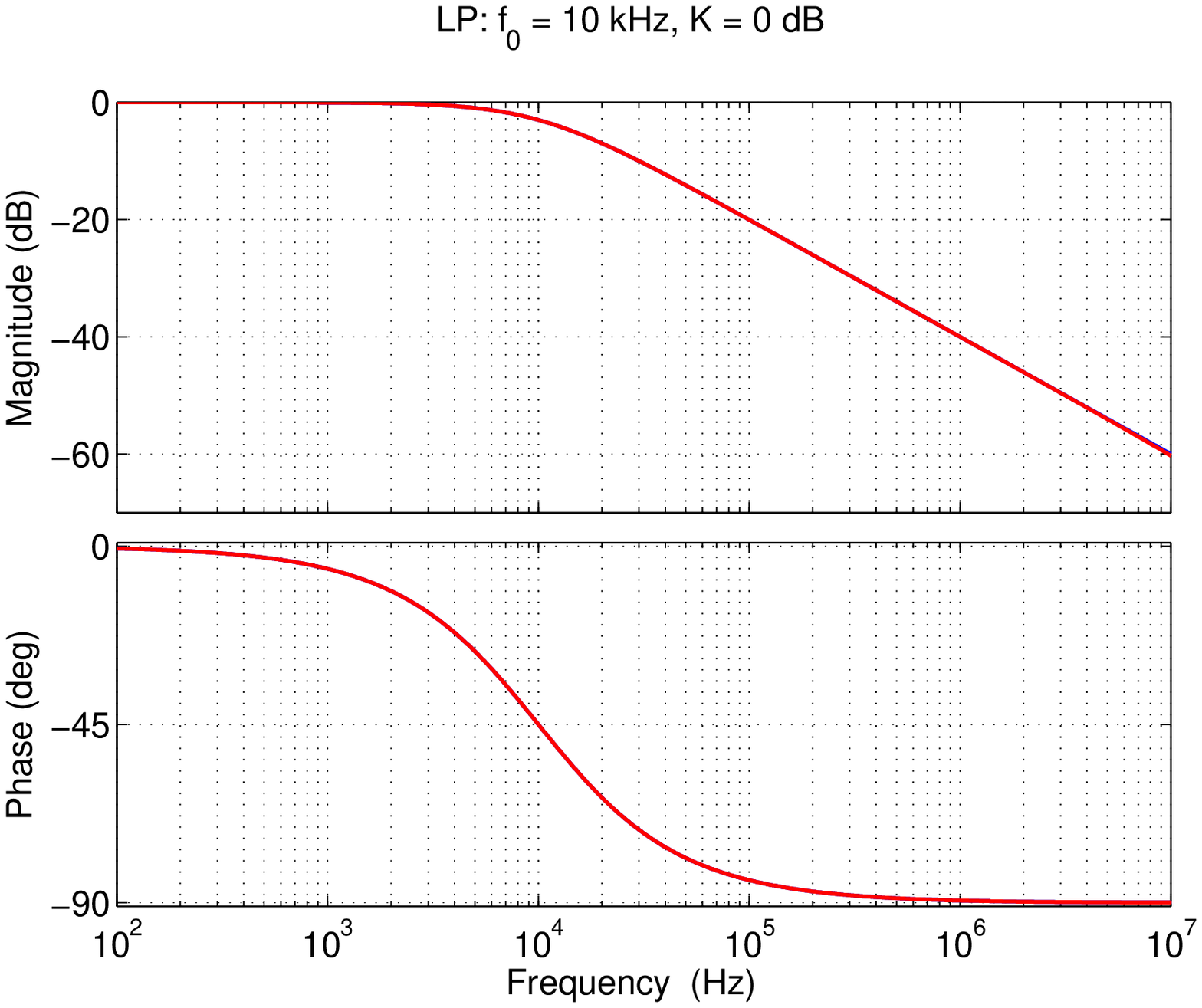} & \includegraphics[width=1.0\columnwidth]{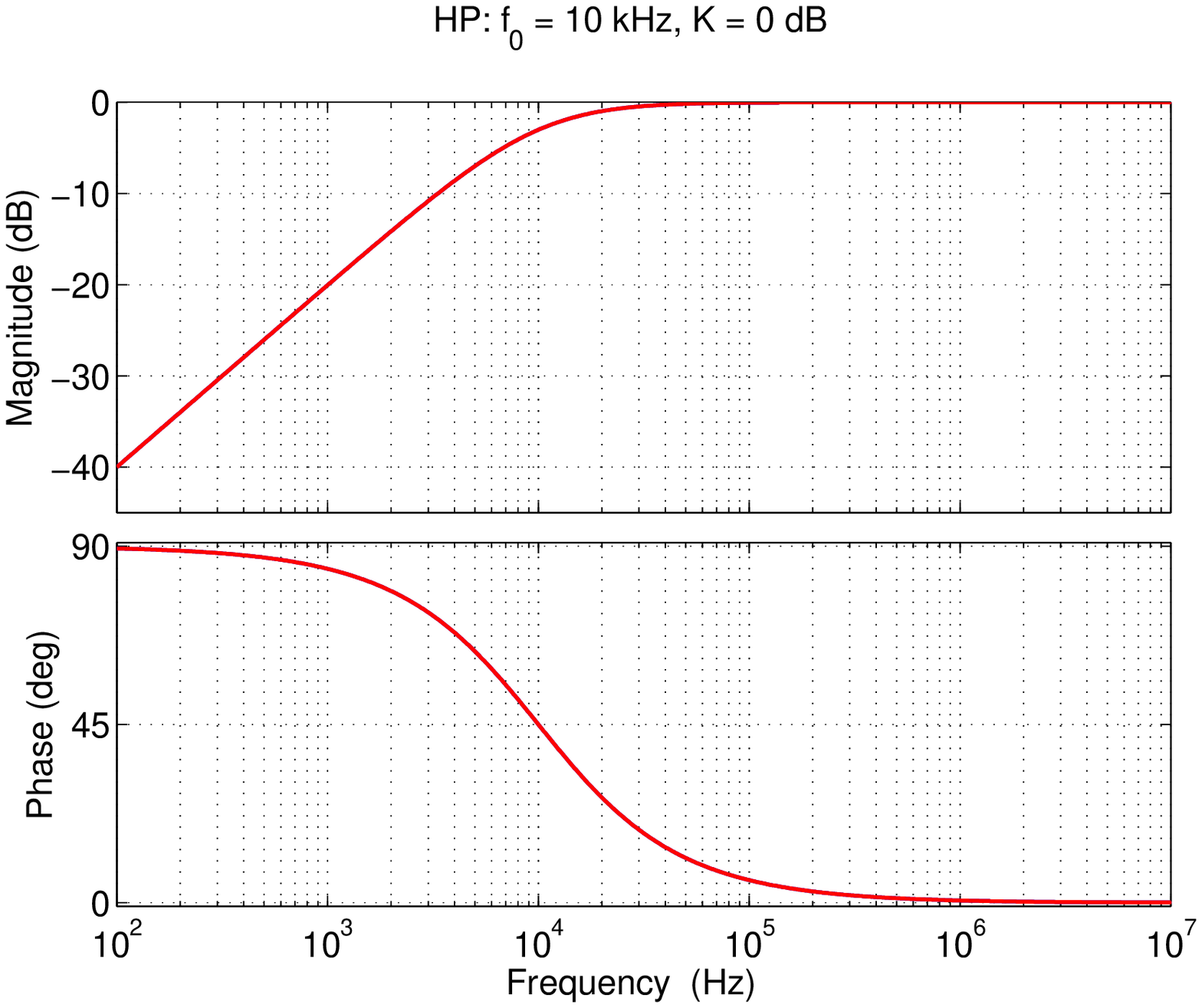} \\
\includegraphics[width=1.0\columnwidth]{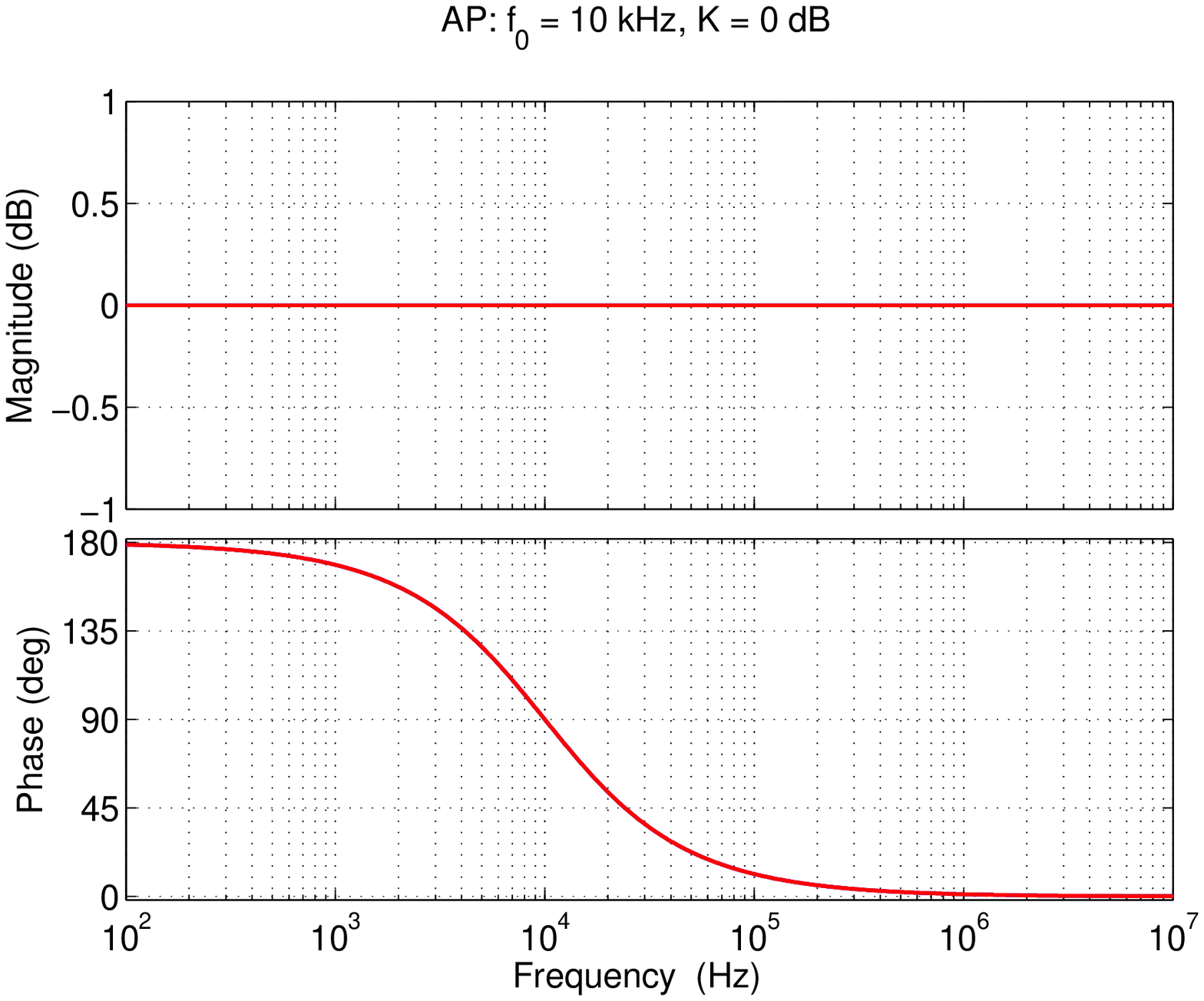} & \includegraphics[width=1.0\columnwidth]{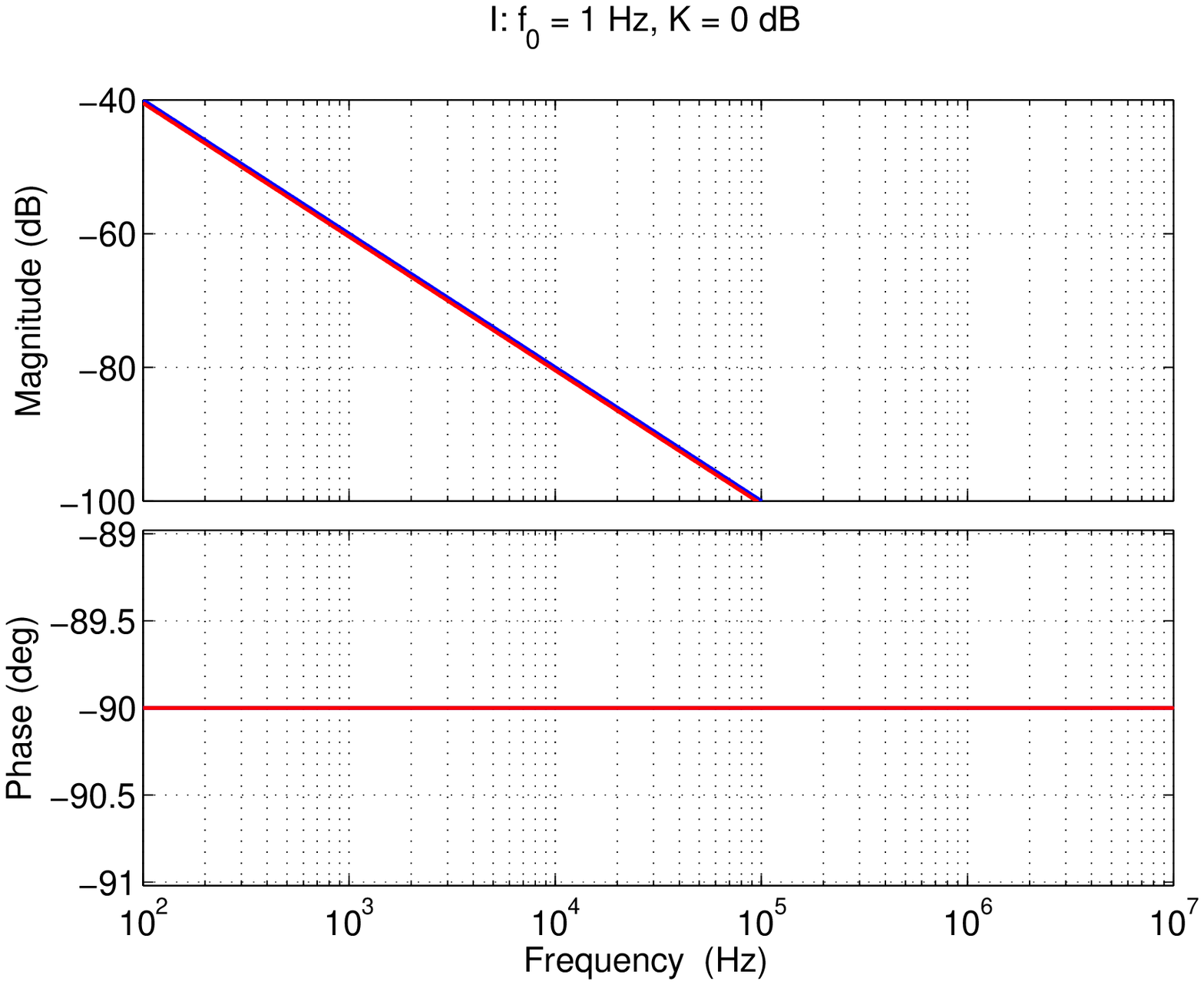} \\
\includegraphics[width=1.0\columnwidth]{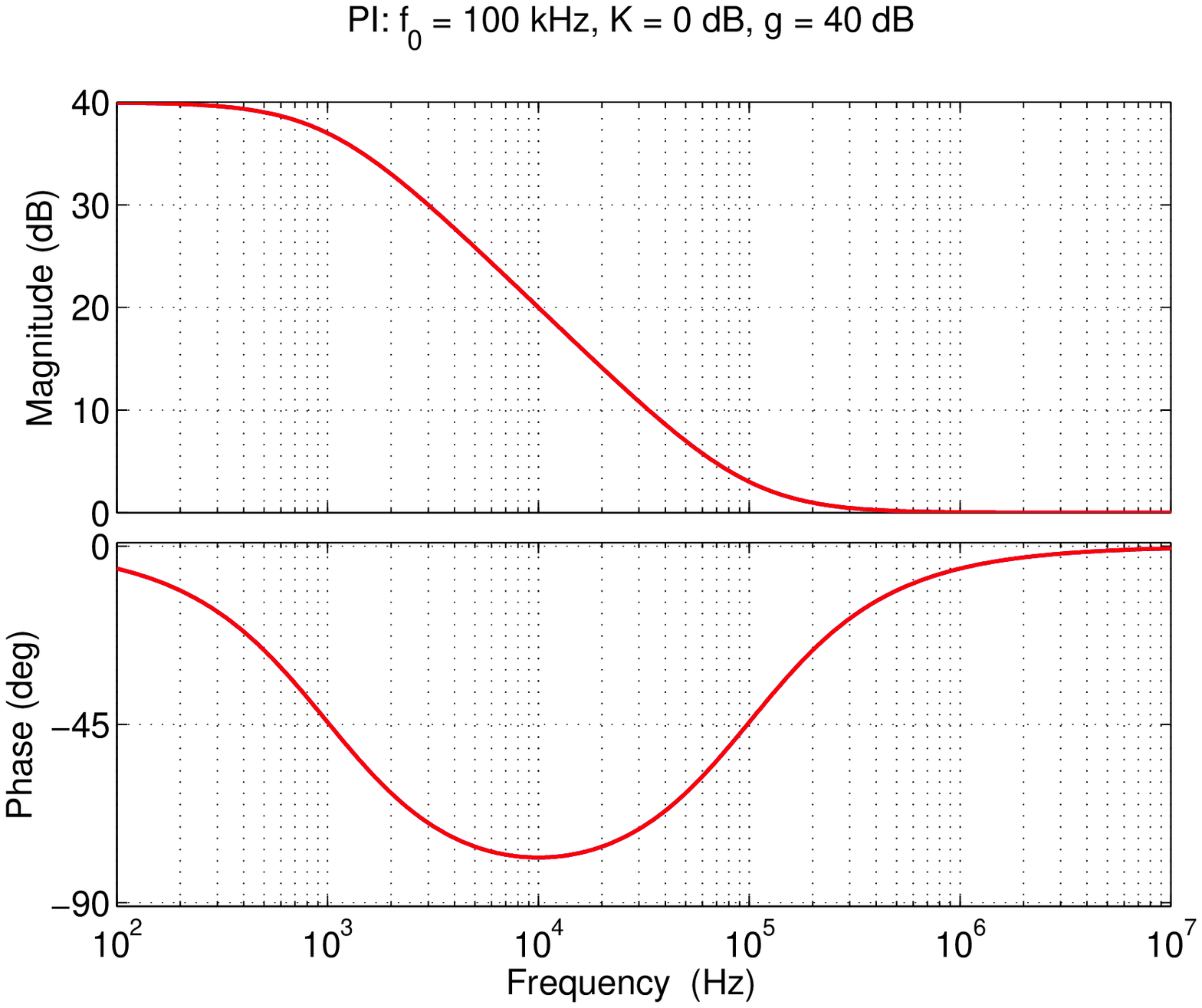} & \includegraphics[width=1.0\columnwidth]{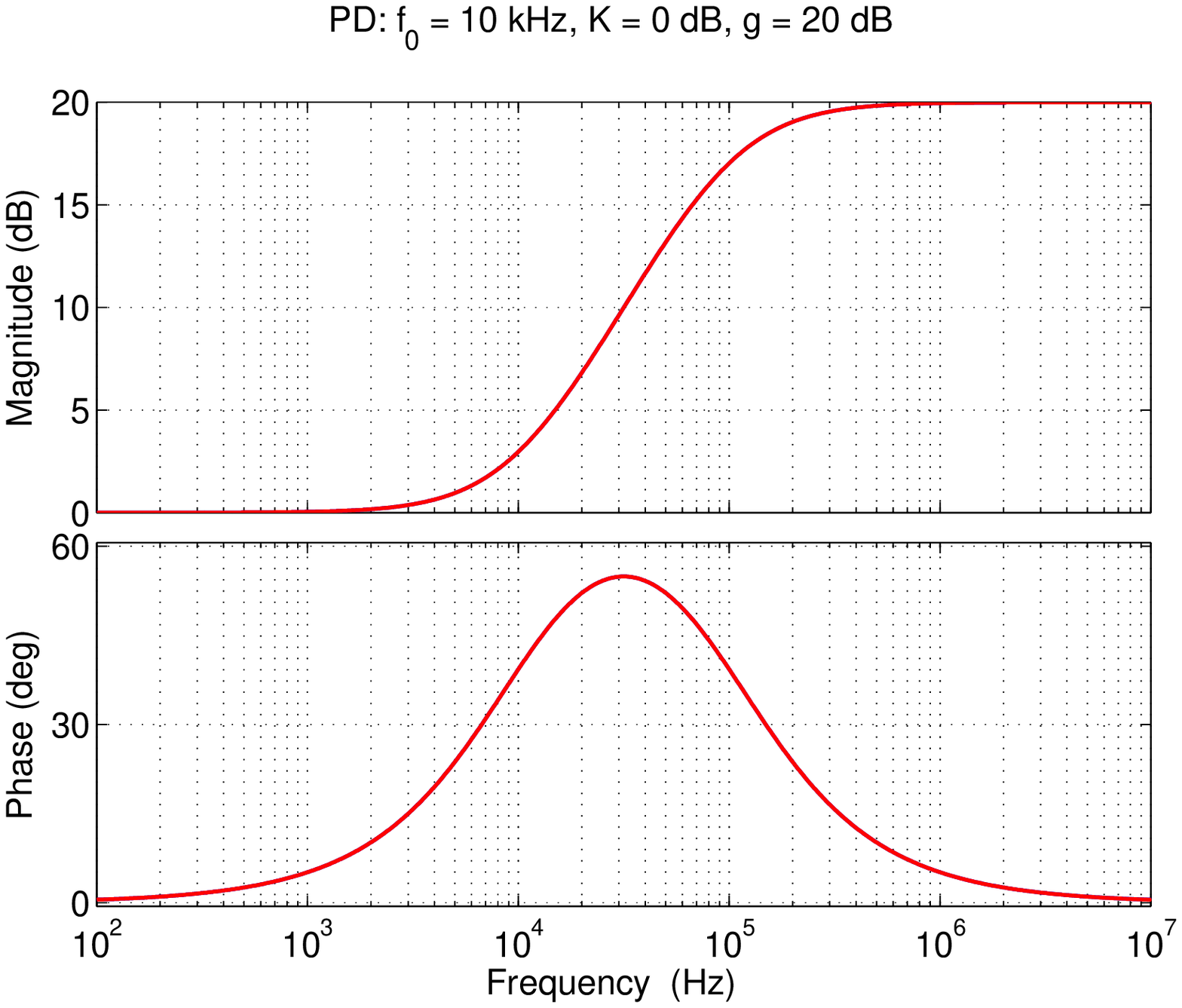}
\end{array}
$
\caption{\label{fig:IIR1}Transfer functions of the first-order IIR filters implemented in the digital servo.  The continuous time, analytic transfer function is plotted in blue, and the discrete time, integer math transfer function is plotted in red.  Note that the blue curve is covered up by the red curve in most cases, indicating that the discrete transfer function reproduces the continuous transfer function to within the line width.}
\end{center}
\end{figure*}

\begin{figure*}
\begin{center}
$
\begin{array}{cc}
\includegraphics[width=1.0\columnwidth]{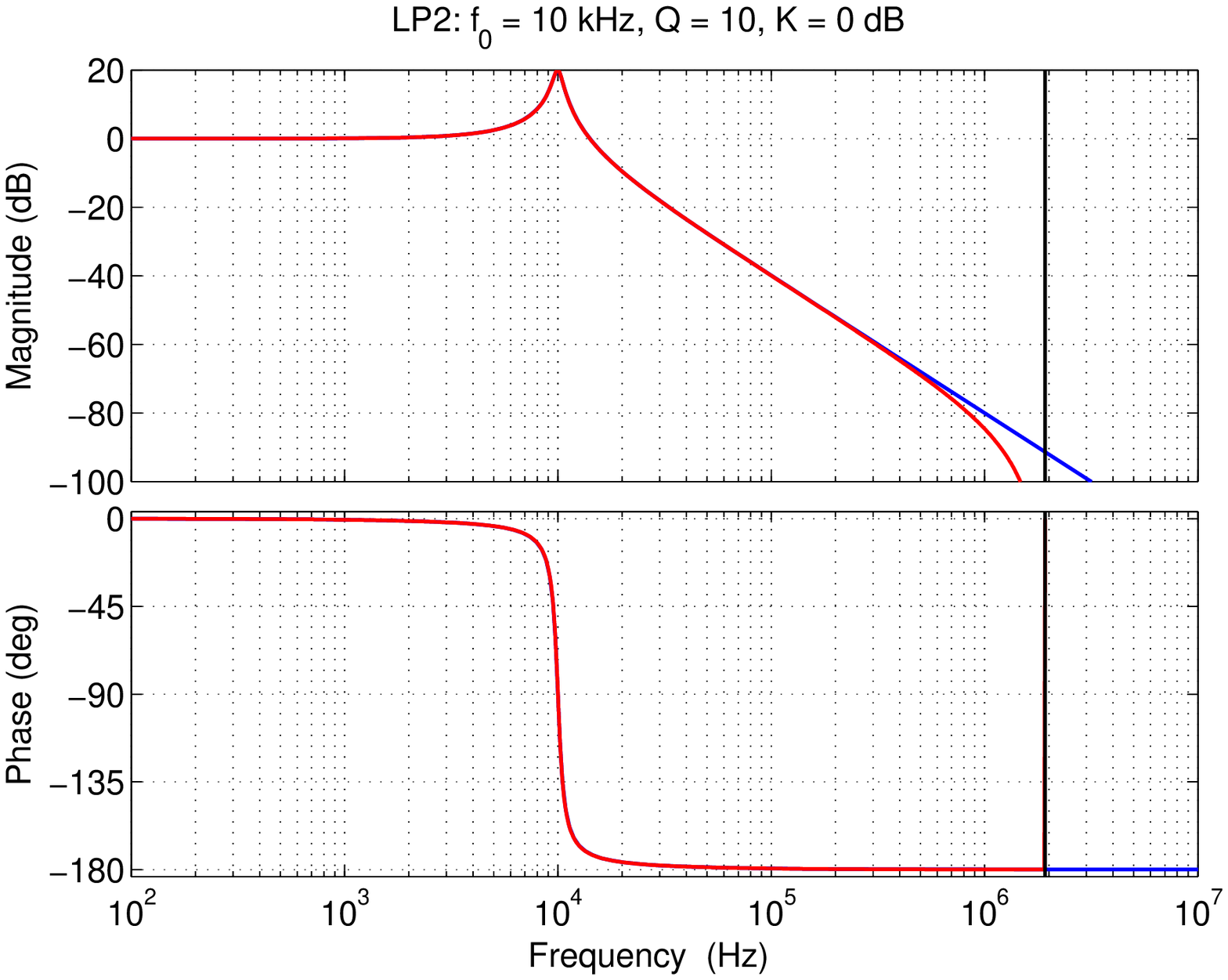} & \includegraphics[width=1.0\columnwidth]{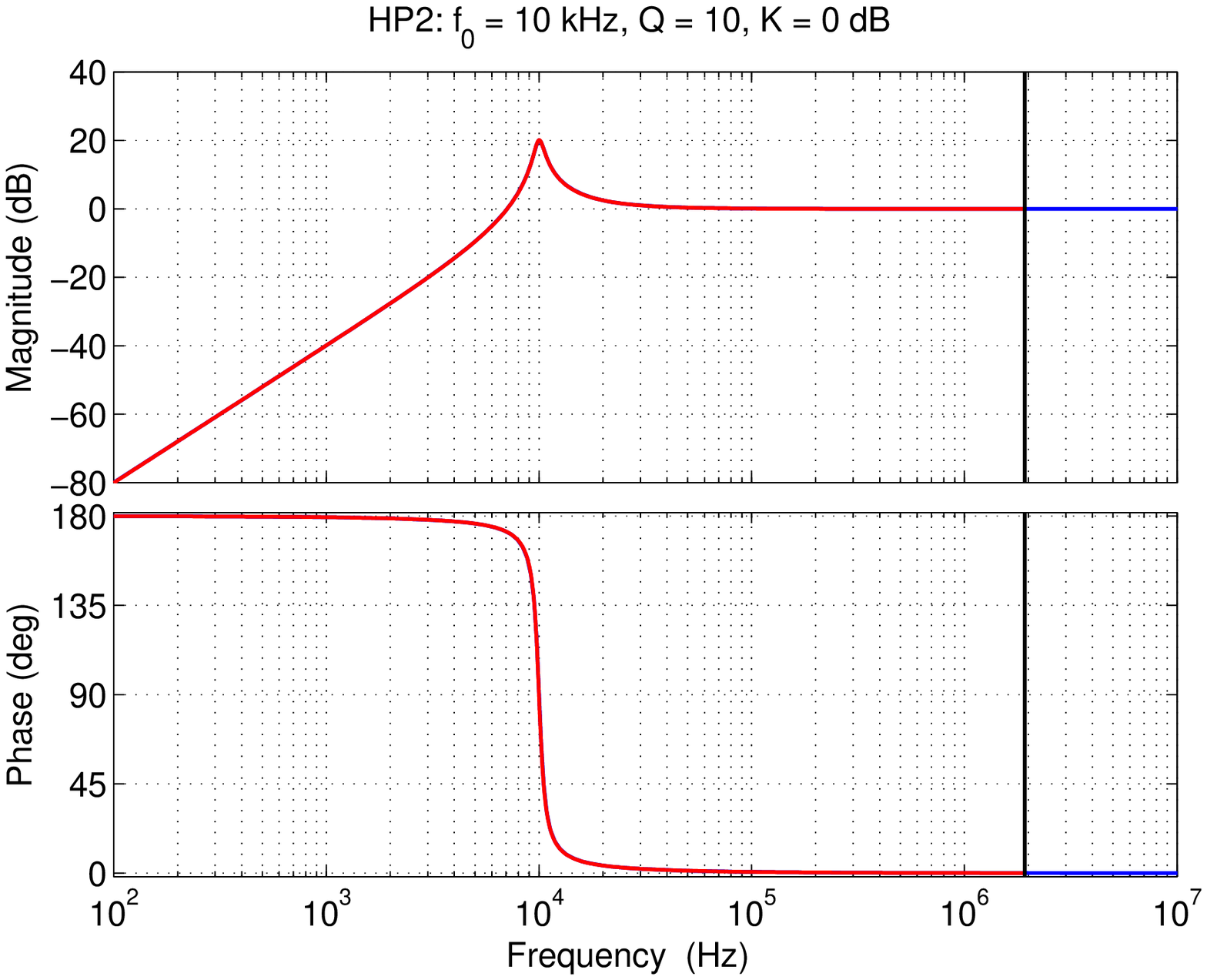} \\
\includegraphics[width=1.0\columnwidth]{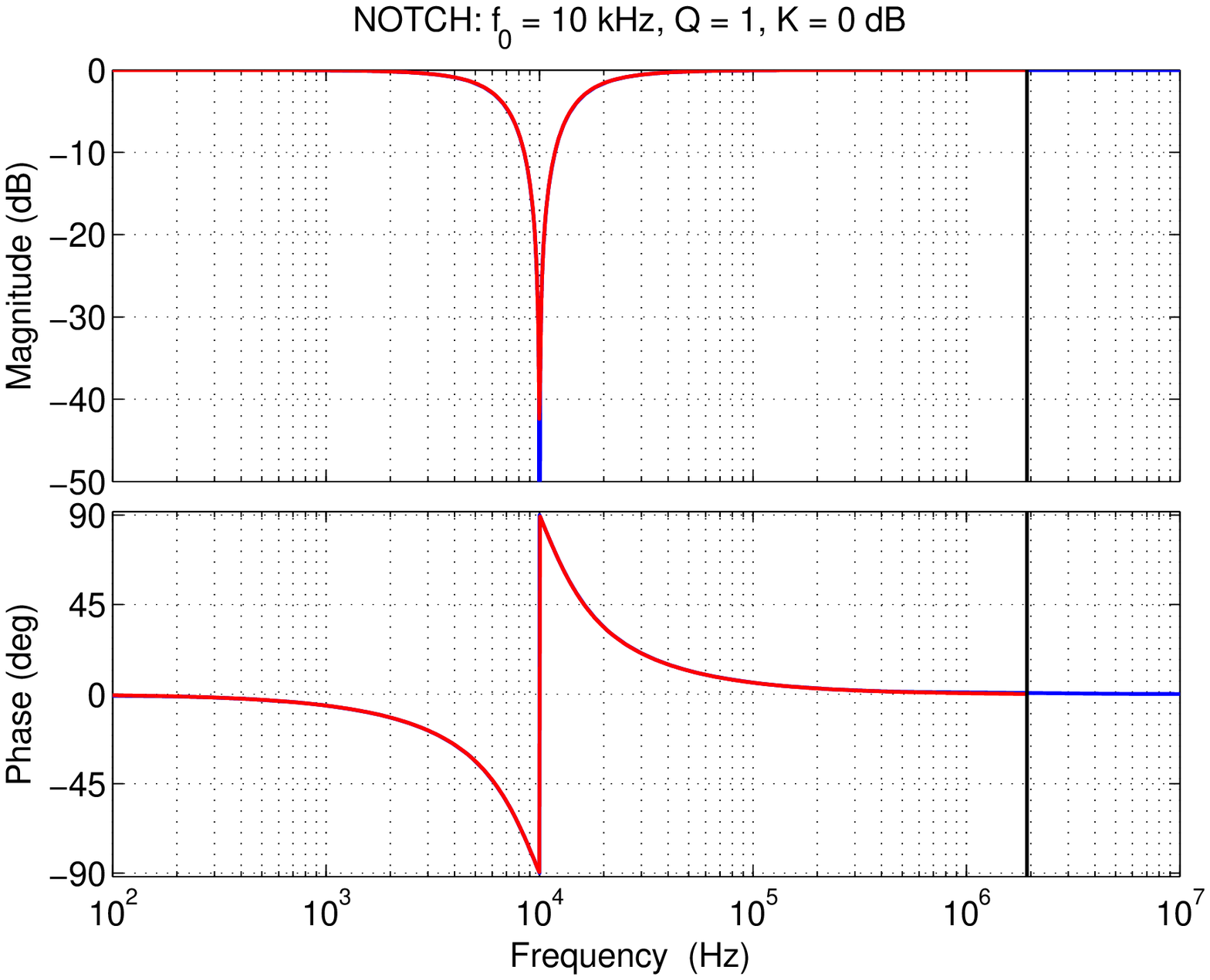} & \includegraphics[width=1.0\columnwidth]{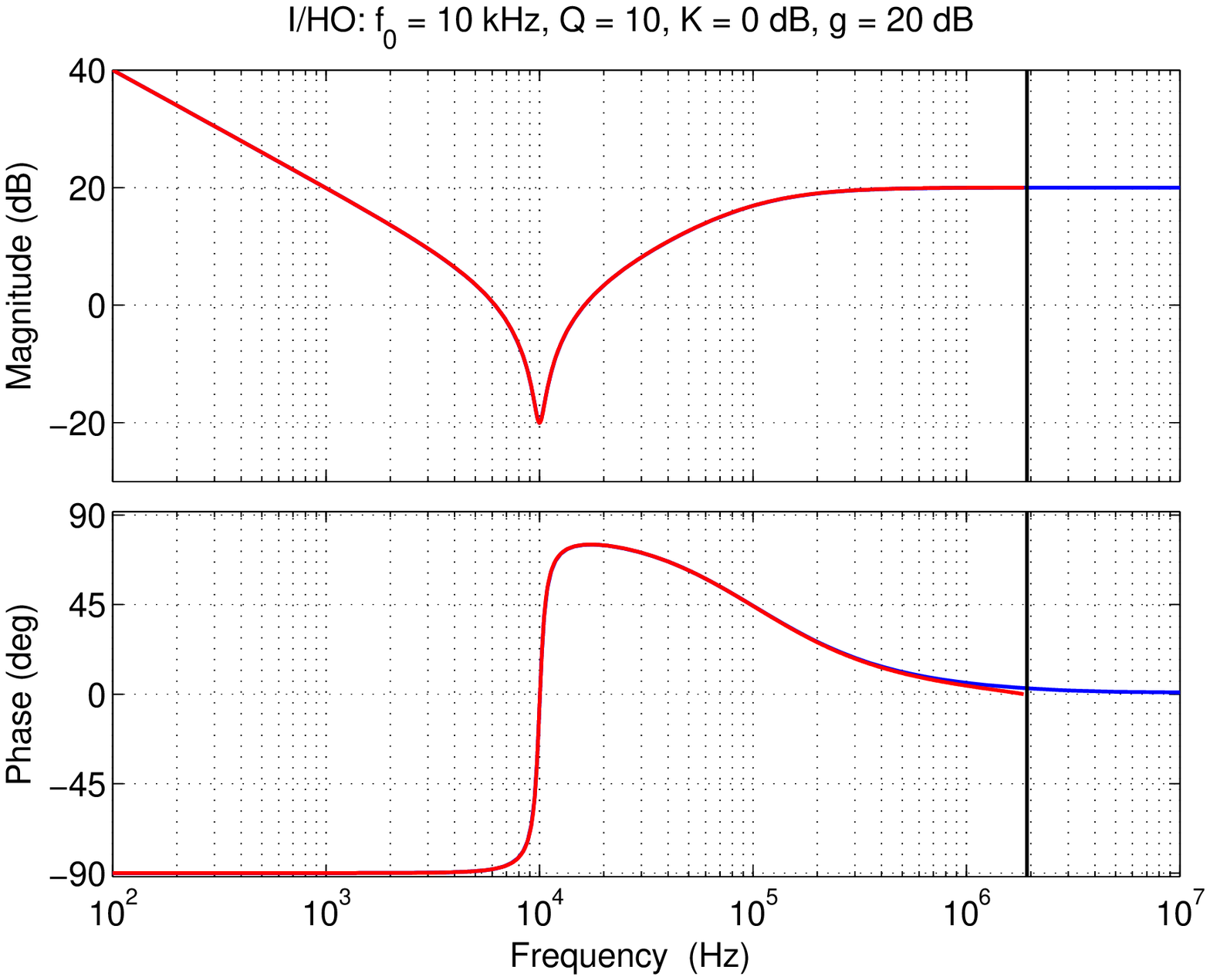}
\end{array}
$
\caption{\label{fig:IIR2}Transfer functions of the second-order IIR filters implemented in the digital servo.  The continuous time, analytic transfer function is plotted in blue, and the discrete time, integer math transfer function is plotted in red.  The vertical black line denotes the Nyquist frequency of the IIR filter update rate.}
\end{center}
\end{figure*}

Here, we design IIR filters by first writing down a continuous transfer function, second converting to a discrete transfer function, and third extracting the corresponding IIR filter.  These three steps are described below for one example: a gain-limited PI filter.  Table \ref{tab:IIR} and Figs.~\ref{fig:IIR1} and \ref{fig:IIR2} catalog the continuous and discrete versions of the first- and second-order filters implemented in the digital servo.

\begin{turnpage}
\begin{table*}
\caption{\label{tab:IIR}Catalog of IIR filters implemented in the digital servo.  Note that the PI and PD filters are gain-limited such that the maximum gain is $K g$, and that we have defined $\tilde{f}_0 \equiv \pi f_0 T_s$.  The parameter ranges are selected such that rounding errors do not cause the discretized transfer function to be significantly different from the continuous transfer function.}
\begin{ruledtabular}
{\renewcommand{\arraystretch}{2.5}
\begin{tabular}{rlllllllllll}
Type		& Continuous 		& \multicolumn{5}{l}{IIR filter coefficients}									&\multicolumn{5}{l}{Scaling and parameter ranges} \\
		& transfer function		& $a_1/a_0$		& $a_2/a_0$		& $b_0/a_0$		& $b_1/a_0$		& $b_2/a_0$		& $a_0$		& $f_0$ [Hz]		& $Q$		& $K$ [dB]		& $g$ [dB] \\
\hline
LP		& $\frac{K}{1 + \frac{s}{2 \pi f_0}}$							& $\frac{1 - \tilde{f}_0}{1 + \tilde{f}_0}$		&& $\frac{K \tilde{f}_0}{1 + \tilde{f}_0}$		& $\frac{K \tilde{f}_0}{1 + \tilde{f}_0}$		&& $2^{26}$	& $[1, 10^7]$	&& $[0, 40]$ \\
HP		& $\frac{K}{1 + \frac{2 \pi f_0}{s}}$							& $\frac{1 - \tilde{f}_0}{1 + \tilde{f}_0}$		&& $\frac{K}{1 + \tilde{f}_0}$				& $- \frac{K}{1 + \tilde{f}_0}$				&& $2^{26}$	& $[1, 10^7]$	&& $[-40, 40]$ \\
AP		& $K \frac{\frac{s}{2 \pi f_0} - 1}{\frac{s}{2 \pi f_0} + 1}$				& $\frac{1 - \tilde{f}_0}{1 + \tilde{f}_0}$		&& $K \frac{1 - \tilde{f}_0}{1 + \tilde{f}_0}$		& $-K$							&& $2^{26}$	& $[1, 10^7]$	&& $[0, 40]$ \\
I		& $K \frac{2 \pi f_0}{s}$									& $1$								&& $K \tilde{f}_0$						&  $K \tilde{f}_0$						&& $2^{26}$	& $[1]$		&& $[0, 200]$ \\
PI		& $K \frac{1 + \frac{s}{2 \pi f_0}}{\frac{1}{g} + \frac{s}{2 \pi f_0}}$		& $\frac{1 - \tilde{f}_0/g}{1 + \tilde{f}_0/g}$		&& $K \frac{1 + \tilde{f}_0}{1 + \tilde{f}_0/g}$	& $- K \frac{1 - \tilde{f}_0}{1 + \tilde{f}_0/g}$	&& $2^{26}$	& $[10, 10^6]$	&& $[-40, 40]$		& $[5, \infty]$ \\
P		& $K$												& $0$								&& $K$							& $0$								&& $2^{26}$	&			&& $[-200, 200]$ \\
PD		& $K \frac{1 + \frac{s}{2 \pi f_0}}{1 + \frac{s}{2 \pi f_0 g}}$			& $\frac{1/g - \tilde{f}_0}{1/g + \tilde{f}_0}$		&& $K \frac{1 + \tilde{f}_0}{1/g + \tilde{f}_0}$	& $- K \frac{1 - \tilde{f}_0}{1/g + \tilde{f}_0}$	&& $2^{26}$	& $[10, 10^6]$	&& $[-40, 0]$		& $[5, 30]$ \\
\hline
LP2		& $\frac{K}{1 + \frac{s}{2 \pi f_0 Q} + \left(\frac{s}{2 \pi f_0}\right)^2}$
		& $\frac{2 \left[1 - \tilde{f}_0^2 \right]}{1 + \frac{\tilde{f}_0}{Q} + \tilde{f}_0^2}$ 			& $- \frac{1 - \frac{\tilde{f}_0}{Q} + \tilde{f}_0^2}{1 + \frac{\tilde{f}_0}{Q} + \tilde{f}_0^2}$
		& $\frac{K \tilde{f}_0^2}{1 + \frac{\tilde{f}_0}{Q} + \tilde{f}_0^2}$					& $\frac{2 K \tilde{f}_0^2}{1 + \frac{\tilde{f}_0}{Q} + \tilde{f}_0^2}$
		& $\frac{K \tilde{f}_0^2}{1 + \frac{\tilde{f}_0}{Q} + \tilde{f}_0^2}$
		& $2^{32}$		& $[10^2, 10^6]$		& $[0.5, 10^2]$		& $[0]$ \\
HP2		& $\frac{K}{1 + \frac{2 \pi f_0}{s Q} + \left(\frac{2 \pi f_0}{s}\right)^2}$
		& $\frac{2 \left[1 - \tilde{f}_0^2 \right]}{1 + \frac{\tilde{f}_0}{Q} + \tilde{f}_0^2}$ 			& $- \frac{1 - \frac{\tilde{f}_0}{Q} + \tilde{f}_0^2}{1 + \frac{\tilde{f}_0}{Q} + \tilde{f}_0^2}$
		& $\frac{K}{1 + \frac{\tilde{f}_0}{Q} + \tilde{f}_0^2}$									& $- \frac{2 K}{1 + \frac{\tilde{f}_0}{Q} + \tilde{f}_0^2}$
		& $\frac{K}{1 + \frac{\tilde{f}_0}{Q} + \tilde{f}_0^2}$
		& $2^{32}$		& $[10^3, 10^5]$		& $[0.5, 10^2]$		& $[0]$ \\
NOTCH	& $\frac{K \left[1 + \left(\frac{s}{2 \pi f_0}\right)^2 \right]}{1 + \frac{s}{2 \pi f_0 Q} + \left(\frac{s}{2 \pi f_0}\right)^2}$
		& $\frac{2 \left[1 - \tilde{f}_0^2 \right]}{1 + \frac{\tilde{f}_0}{Q} + \tilde{f}_0^2}$ 			& $- \frac{1 - \frac{\tilde{f}_0}{Q} + \tilde{f}_0^2}{1 + \frac{\tilde{f}_0}{Q} + \tilde{f}_0^2}$
		& $\frac{K \left[1 + \tilde{f}_0^2 \right]}{1 + \frac{\tilde{f}_0}{Q} + \tilde{f}_0^2}$			& $- \frac{2 K \left[1 - \tilde{f}_0^2\right]}{1 + \frac{\tilde{f}_0}{Q} + \tilde{f}_0^2}$
		& $\frac{K \left[1 + \tilde{f}_0^2 \right]}{1 + \frac{\tilde{f}_0}{Q} + \tilde{f}_0^2}$
		& $2^{32}$		& $[10^2, 10^6]$		& $[0.5, 10]$		& $[0]$ \\
\hline
I/HO		& $\frac{K}{1 + \frac{s}{2 \pi f_0 g}} \left[\frac{2 \pi f_0}{s} + \frac{1}{Q} + \frac{s}{2 \pi f_0} \right]$
		& $\frac{2}{1 + \tilde{f}_0 g}$															& $-\frac{1 - \tilde{f}_0 g}{1 + \tilde{f}_0 g}$
		& $K \frac{1 + \frac{\tilde{f}_0}{Q} + \tilde{f}_0^2}{\frac{1}{g} + \tilde{f}_0}$					& $-2 K \frac{1 - \tilde{f}_0^2}{\frac{1}{g} + \tilde{f}_0}$
		& $K \frac{1 - \frac{\tilde{f}_0}{Q} + \tilde{f}_0^2}{\frac{1}{g} + \tilde{f}_0}$
		& $2^{26}$		& $[10^2, 10^5]$		& $[10^{-2}, 10^2]$	& $[0]$	& $[20, 40]$ \\
\end{tabular}
}
\end{ruledtabular}
\end{table*}
\end{turnpage}

\subsection{Continuous transfer function}

Formally, the transfer function is obtained by Laplace transformation of the equations of motion of the dynamical system.  In practice, however, controller transfer function design typically proceeds in the reverse direction: first a transfer function is postulated and second a corresponding controller (i.e., dynamical system) is built.  For a gain-limited PI filter, we postulate the transfer function
\begin{equation}
H(s) = K \frac{1 + \frac{s}{2 \pi f_0}}{\frac{1}{g} + \frac{s}{2 \pi f_0}}
\end{equation}
where $K$ is the proportional gain, $f_0$ is the PI corner frequency, and $g$ is the gain limit.  To convert from the Laplace transformation variable $s$ to angular frequency $\omega$, simply evaluate $H(s)$ at $s = \textrm{i} \omega$.

\subsection{Discrete transfer function}

To go from a continuous transfer function $H(s)$ to a discrete transfer function $H(z)$, we use Tustin's transformation (an approximate version of the $z$ transformation):
\begin{equation}
s \rightarrow \frac{2}{T_s} \frac{1 - z^{-1}}{1 + z^{-1}} \ ,
\end{equation}
where $T_s$ is the sample time.  This corresponds to discretization of the equations of motion using the trapezoidal rule for integration.  For the case of the gain-limited PI filter, we get
\begin{equation}
H(z) = \frac{K \frac{1 + \pi f_0 T_s}{1 + \pi f_0 T_s/g} - K \frac{1 - \pi f_0 T_s}{1 + \pi f_0 T_s/g} z^{-1}}
{1 - \frac{1 - \pi f_0 T_s/g}{1 + \pi f_0 T_s/g} z^{-1}} \ .
\end{equation}

\subsection{IIR filter}

The inverse of the $z$ transformation variable, $z^{-1}$, means ``delay by $T_s$''.  As a result, the general discrete transfer function
\begin{equation}
H(z) = \frac{b_0 + b_1 z^{-1} + b_2 z^{-2} + \cdots}{a_0 - a_1 z^{-1} - a_2 z^{-2} - \cdots}
\end{equation}
can be interpreted as an IIR filter with a corresponding difference equation
\begin{eqnarray}\label{eq:IIRdifferenceEqn}
y_n &=& \left( a_1 y_{n-1} + a_2 y_{n-2} + \cdots \right. \nonumber \\
	&& \left. + \ b_0 x_n + b_1 x_{n-1} + b_2 x_{n-2} + \cdots \right) / a_0 \ ,
\end{eqnarray}
where $x_i$ are the inputs and $y_i$ are the outputs.  For the case  of the gain-limited PI filter, the IIR filter coefficients are given by
\begin{eqnarray}
a_0 & = & 1 \\
a_1 & = & \frac{1 - \pi f_0 T_s/g}{1 + \pi f_0 T_s/g} \\
b_0 & = & K \frac{1 + \pi f_0 T_s}{1 + \pi f_0 T_s/g} \\
b_1 & = & - K \frac{1 - \pi f_0 T_s}{1 + \pi f_0 T_s/g} \ .
\end{eqnarray}

\subsection{Implementation details}

\begin{figure*}
\begin{center}
\includegraphics[width=0.9\textwidth]{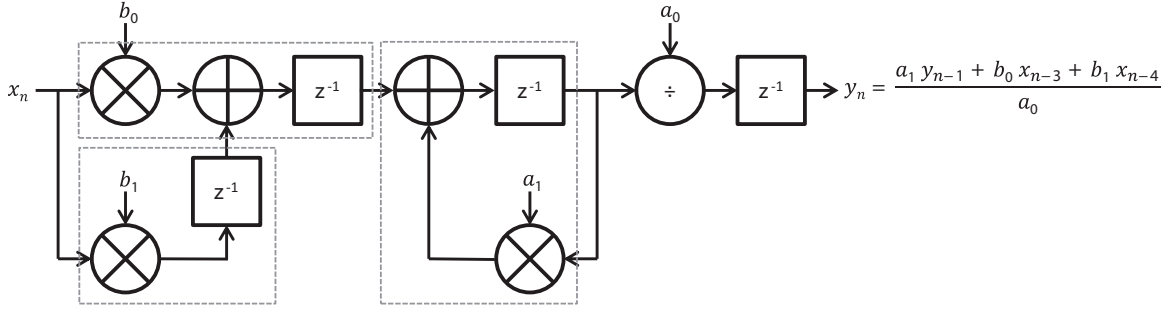}
\caption{\label{fig:IIRfilterImplementation}Block diagram showing details of the implementation of the first-order IIR filters used in the digital servo.  The blocks labeled $z^{-1}$ denote registers which delay the signal by one time step.  Each dashed gray box is a $35 \times 35$ multiply-add operation that uses four combined DSP48A1 slices, followed by an output register.  The division by $a_0$ is actually a bit shift operation and does not require any DSP48A1 slices.  The latency is three clock cycles.}
\end{center}
\end{figure*}

A final consideration is that on the FPGA, IIR filters are implemented with finite precision (fixed point) math.  Thus, it is important to consider truncation of both the filter coefficients and the data, as well as potential overflow due to high gain.  In our implementation, the gain-limited PI filter coefficients above are scaled by $2^{26}$ and we use 35 bit $\times$ 35 bit signed integer multiplication.  Figs.~\ref{fig:IIR1} and \ref{fig:IIR2} show the finite precision transfer functions of the IIR filters built in to the digital servo; their good agreement with the continuous transfer functions indicates that we are using enough bits.

The first-order IIR filters are implemented in the FPGA as shown in Fig.~\ref{fig:IIRfilterImplementation}.  A new output is calculated every clock cycle, and the latency is three clock cycles.  The second-order IIR filters are implemented using a finite state machine that calculates a new output every 27 clock cycles based on Eq.~\ref{eq:IIRdifferenceEqn}.  A single $35 \times 35$ bit multiplier is used in a time multiplexed fashion to sequentially compute all five of the multiplications, and the latency is eight clock cycles.

\end{document}